\documentstyle[preprint,aps,pra]{revtex}
\begin{document} 
\draft
\title {{\em Ab-initio} simulation of high-temperature liquid selenium}
\author{F. Kirchhoff, M. J. Gillan and J.  M.  Holender}
\address{ Physics Department, Keele University \\
Keele, Staffordshire ST5 5BG, U.K.}
\date{\today}
\maketitle
\begin{abstract}
{\em Ab initio} molecular dynamics simulation is used to investigate the
structure and dynamics of liquid Se at temperatures of
870 and 1370~K. The calculated static structure factor is in
excellent agreement with experimental data. The calculated
radial distribution function gives a mean coordination number close to 2,
but we find a significant
fraction of one-fold and three-fold atoms, particularly
at 1370~K, so that the chain structure is considerably disrupted.
The self-diffusion coefficient has values
($\sim 1 \times 10^{-8}$~m~s$^{-1}$) typical of liquid metals.
\end{abstract}

\section{Introduction}

Liquid Se is a highly unusual liquid because of the strong temperature
dependence of its electrical conductivity, self-diffusion coefficient,
viscosity, and other properties. In the trigonal crystal structure
that is stable under ambient conditions, the Se atoms are bonded
into infinite chains, with a rather weak interaction between
chains. Diffraction measurements indicate two-fold coordination in the
liquid, and the persistence of the chain structure in the liquid state
is believed to account for the high viscosity and low
diffusion coefficient at temperatures near the melting point (494~K).
As the temperature is increased to over 1000~K, the viscosity drops
by more that an order of magnitude\cite{glazov}, and the diffusion
coefficient increases by a factor of $\sim$10\cite{axm70}.
Nevertheless, the mean coordination number deduced from diffraction
measurements remains close to 2.0 even for temperatures
well above 1000~K. It seems likely that the chain structure
is strongly disrupted at high temperatures, but it is difficult to
be certain of this from diffraction data alone.
The present paper reports {\em ab initio} molecular dynamics
(AIMD) simulations of high-temperature $\ell$-Se, which shed light
on the three-dimensional structure of the liquid. In addition,
we present information which is relevant to the study of
liquid Se-based alloys such as Ag-Se and Ga-Se, for which we
report AIMD results elsewhere in these proceedings \cite{agse,gase}.

In the AIMD technique, the energy of the system and the forces on
the atoms are obtained from an {\em ab initio} determination
of the electronic ground state for every instantaneous set of
atomic positions. No assumptions are made about the interactions
between atoms, and the technique is expected to give an accurate
description of the covalent bonding present in any state of
thermal equilibrium. The technique is therefore fundamentally
different from the more traditional types of MD simulation
based on assumed interatomic potentials. AIMD simulation has
been used to study many liquid metals and semiconductors, and its
accuracy and reliability are now well established. General reviews
of the technique can be found in refs. \cite{df}.

\section{Simulation Methods}
\label{sec:comput}

Our AIMD simulations are based on {\em ab initio}
molecular dynamics (AIMD),
using density functional theory within the local density
approximation.  We use a plane-wave basis to expand the
valence orbitals, and the valence-core interaction is
represented by a pseudopotential.
At every step of the simulation, we solve the one-electron Kohn-Sham
equations self-consistently. This is done by minimizing the total energy
of the system using a conjugate-gradient technique \cite{pay92}. At every
MD step, the total energy was converged to better than
$2 \times 10^{-5}$~eV
per atom. From the knowledge of the ground state, the forces on the atoms
are calculated using the Hellmann-Feynman theorem. The forces are then
used to integrate the classical equations of motion of the ions.
To handle the semi-metallic nature of the system, which we expect at
the temperatures and pressures we consider here, we use
Fermi-surface smearing, with the electronic occupation numbers treated as
auxiliary dynamical variables \cite{gil89,kre94,gru94,jmh95}.

We use {\em ab-initio} norm-conserving non-local pseudopotentials.
The exchange and correlation energy is included {\em via} the local density
approximation in Ceperley-Alder form\cite{lda}.
The pseudopotential for Se was constructed using the standard Kerker
method \cite{ker}.  The $s$ and $p$ components of the
pseudopotential were generated using the atomic configuration
4$s^{2}$4$p^{4}$, and the $d$ component using the
configuration 4$s^{2}$4$p^{2.75}$3$d^{0.25}$.
The core radii were chosen to be 2.0, 2.0, 2.3 a.u for $s$, $p$ and
$d$ components respectively.
We use the pseudopotential in Kleinman-Bylander separable form \cite{kb}
with the $p$-component treated as local;
the non-local parts of the pseudopotential are treated in real
space \cite{kin91}.

The system we simulated was composed of 69 atoms in a cubic
cell with periodic boundary conditions. The wavefunctions at the
$\Gamma$-point were expanded in a plane wave basis with a cut-off
energy of 11 Ry. Test calculations done on crystalline selenium
($c$-Se) have shown that this cut-off is enough to reproduce
the lattice parameters to within a few percent. A Fermi smearing
width of 0.2~eV was used. The integration of the
classical equations of motion
was done by using Verlet's algorithm, with a time step of 3~fs.
The simulations have been performed at 870~K and 1370~K, at densities of
$3.57 \times 10^3$ and $3.04 \times 10^3$~kg~m$^{-3}$.
These densities correspond to experimental
conditions under pressures of 10 and 100 bar respectively for
which neutron\cite{edeling81} and X-ray\cite{tamura91} diffraction measurements
on $\ell$-Se have
been carried out.

In order to avoid long MD runs to equilibrate the system, we generated
initial configurations by performing MD simulations using a simple
empirical tight-binding model \cite{ehmg}. Using this model, the system was
equilibrated for 2.0~ps by rescaling the velocities, and then evolved
microcanonically for another 3.0~ps. The resulting configuration was then
used as initial configuration for the AIMD simulation. The AIMD run at
each temperature had a duration of 3.0\ ps, with averages taken only
over the last 2.0\ ps.

\section{Results and discussion}

In Fig.\ \ref{fig:sf} we compare the structure factors $S(k)$
obtained from our simulations for T = 873 and 1373~K
with the experimental results of neutron diffraction measurements
\cite{edeling81} at the same conditions and X-ray diffraction measurements
\cite{tam92} at T = 873 and 1473~K. The overall
agreement with the experiment is excellent at both temperatures.
The positions and the intensity of the main and the second
peaks, at $\sim$3.6 and $\sim$5.8~\AA$^{-1}$ respectively, are
well reproduced. The same applies for the shoulder
around $\sim$2.3~\AA$^{-1}$
found for the lower temperature. However there is a slight disagreement with
the X-ray data at higher temperature probably due to the different temperature
involved in the experiment (1473K).
The comparison with experiment at low $k$ is more difficult due to the
noise in the calculated $S(k)$ resulting from the small size of our system
and the resulting lack of $k$-space resolution. This could explain the
spurious pre-peak observed just above 1~\AA$^{-1}$ at 1370~K.
We see that with increasing temperature and pressure
the amplitudes of the first and second peak decrease and the
shoulder just above  $\sim$2.3~\AA$^{-1}$ is
damped out.
The positions of the first and the second peak of $S(k)$ are
unaffected by the increasing pressure and temperature.

The pair correlation functions for the two temperatures are displayed
in Fig.\ \ref{fig:rdf} together with the experimental data from
neutron and X-ray diffraction measurements.
The overall agreement with the diffraction data is good. Our calculations
show two prominent peaks for both temperature. Their positions are 2.31 and
3.64 \AA\ at T = 870~K and 2.30 and 3.72 \AA\ at T = 1370~K. These values
directly give the first and second neighbor distances, $r_1$ and $r_2$,
within chains. It should be noted that there is some disagreement among the
available experimental data for the first and second neighbor
distances\cite{tamura91,edeling81}, probably due to the difficulty of deriving
$g(r)$ from $S(k)$.
Nevertheless, our findings are in good agreement with the more recent
X-ray diffraction data \cite{tamura91}, which for
T = 873~K give $r_1 = 2.31$ and $r_2 = 3.65$ and for T = 1473~K
$r_1 = 2.31$ and $r_2 = 3.64$.
We note that there is some disagreement so far as the height of the peaks
is concerned, but
this difference is similar to that observed when experimental results
are compared. The effect of temperature on $g(r)$ is mainly to decrease the
magnitude and increase the width of the the main peak. We also note the
appearance of a shoulder at $\sim$2.75 \AA\ and an increase of $g(r)$ at
the first minimum, $r_c$. The area under the first peak of
$g(r)$ gives the nearest-neighbor coordination number $N_c$. We find that for
both temperatures selenium
has an average coordination number slightly smaller than 2: $N_c$=1.95 at
870~K and 1.98 at 1370~K. This means that the chain structure characteristic
of $c$-Se is preserved even at high temperature,
in agreement with what is found
experimentally \cite{tamura91,edeling81}. The
first-neighbor distance found in
our calculation is shorter than the corresponding
distance in $c$-Se (2.37~\AA),
whereas the second neighbor distance is
larger (3.44~\AA). This implies that
as the interchain distance is increased due to the decreasing density, the
covalent bonds within the chains
contract \cite{tamura91}.

\begin{figure}
\caption{Snapshots of typical configurations of $\ell$-Se
at 1370~K.
Bonds are drawn
between Se atoms with separation $<$ 3.0 \AA. Bonds to atoms in neighboring
cells are represented by two-colored sticks.}
\label{fig:snap2}
\end{figure}
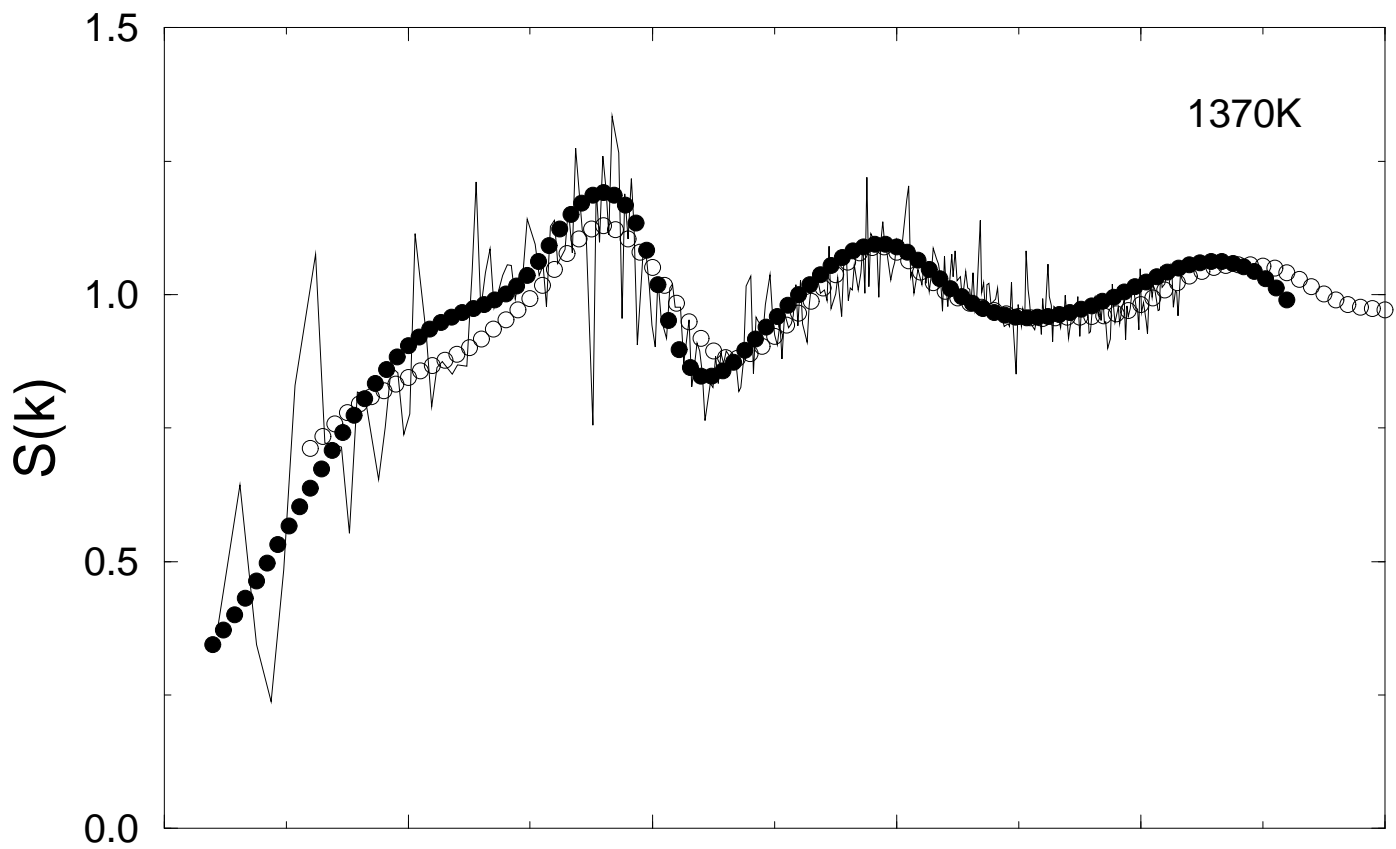
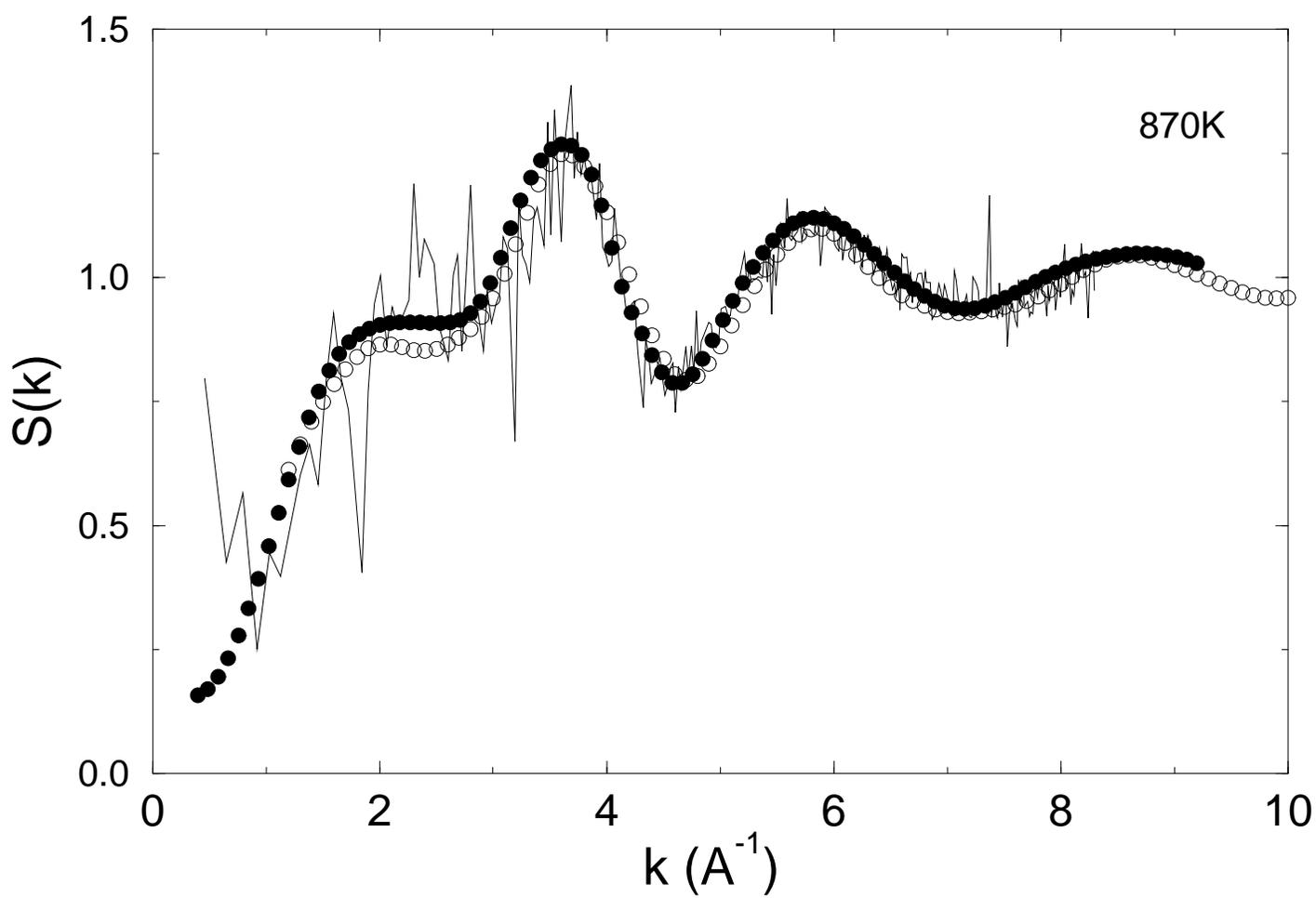
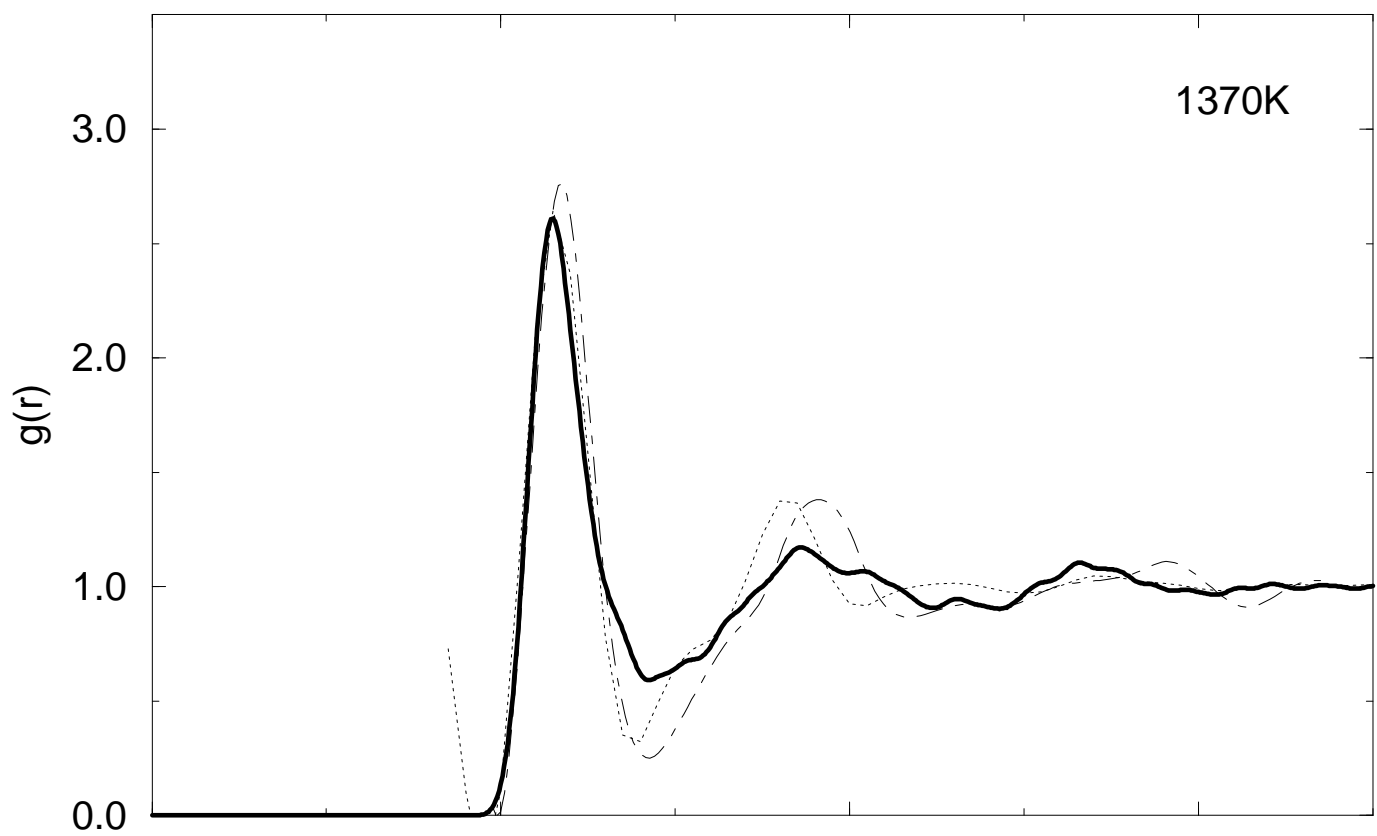
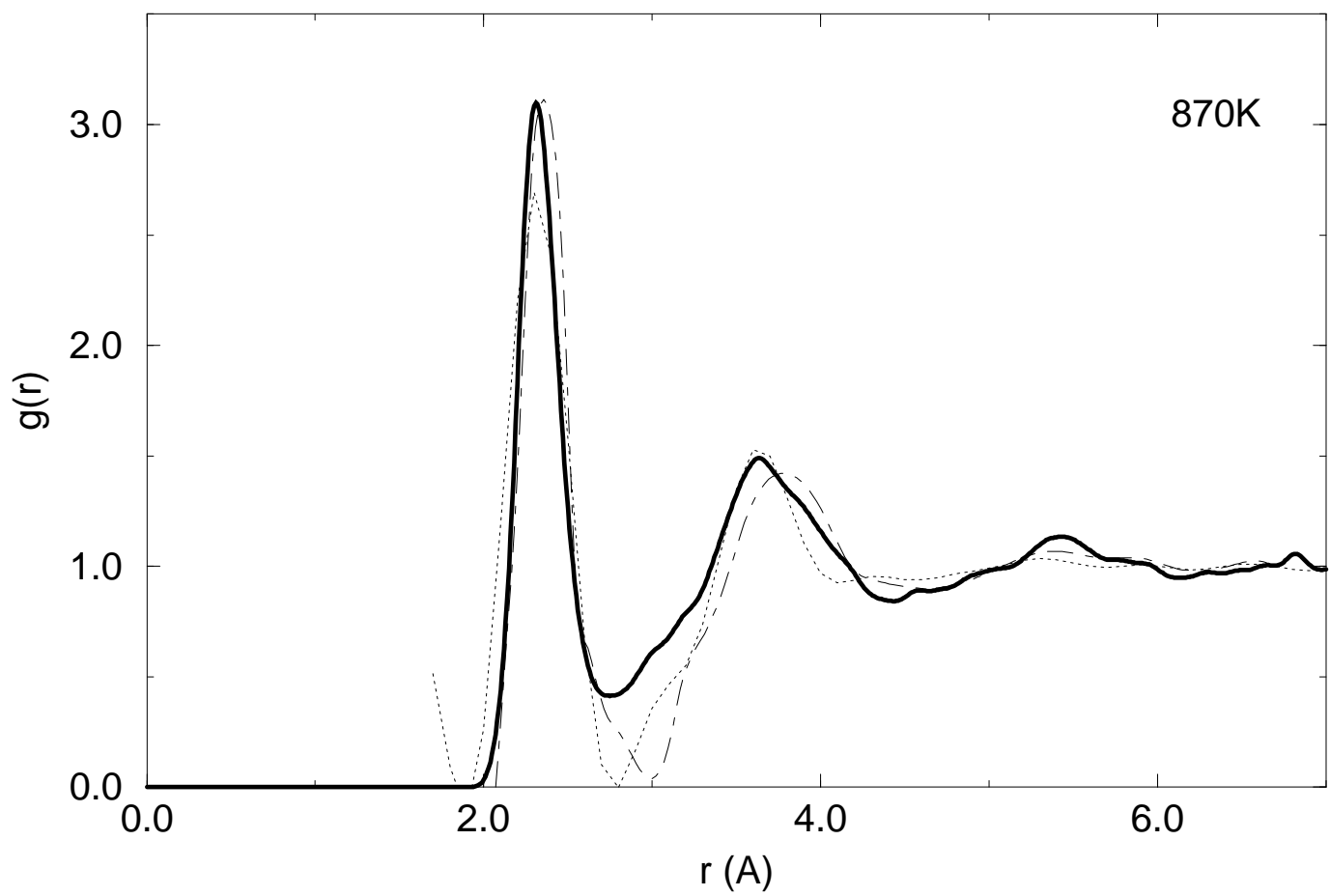
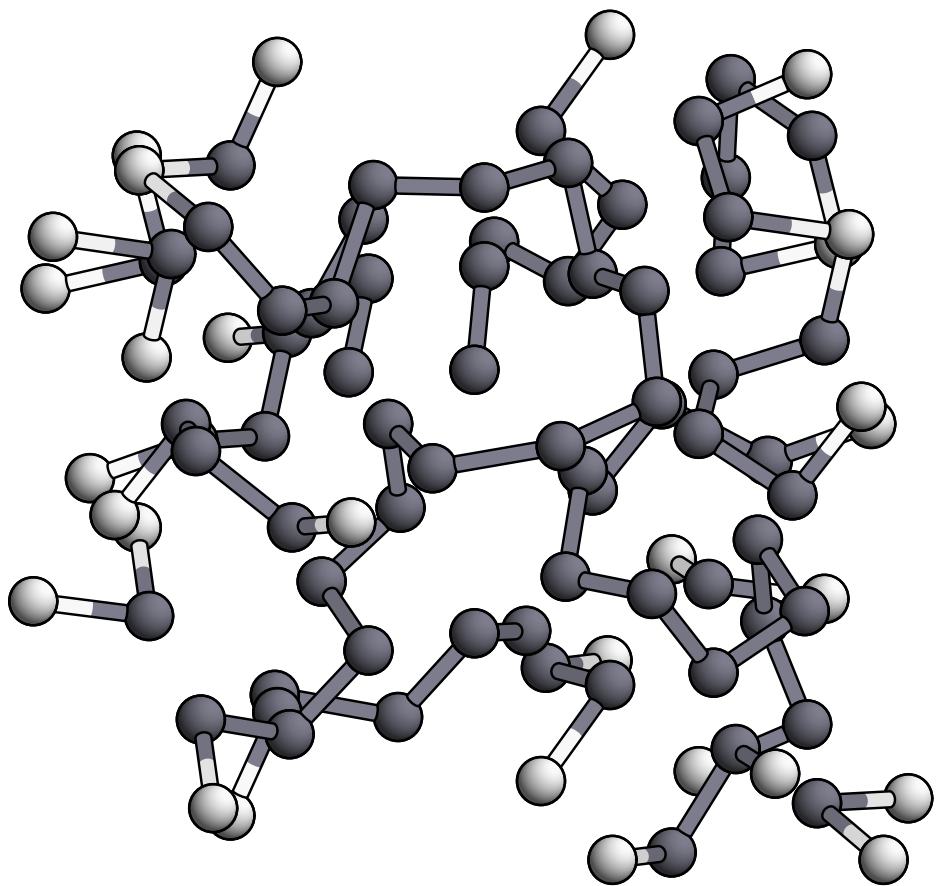
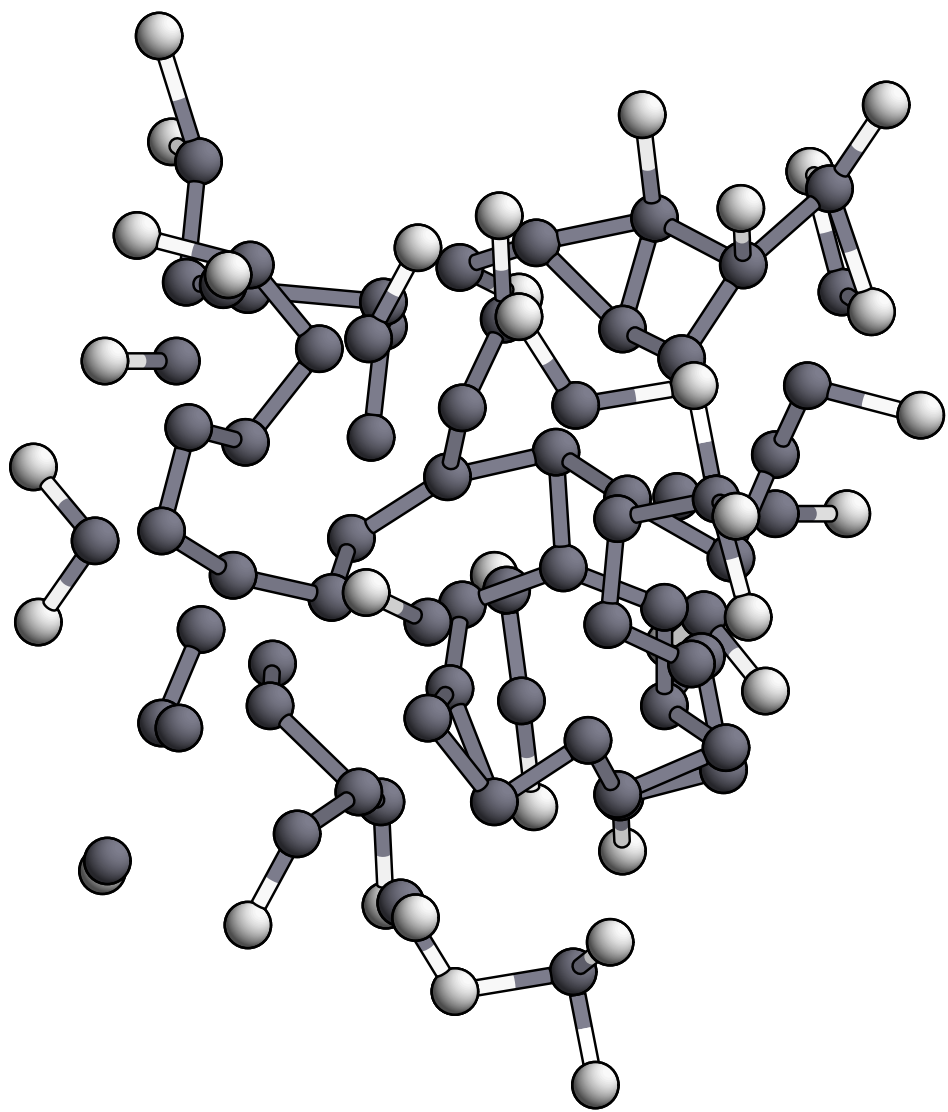

Table \ref{tab:nc}
shows the distribution of first-neighbor coordination numbers found in our
simulations. At both temperatures, the majority of the atoms are two-fold
coordinated but with a significant proportion of one-fold coordinated
atoms, thus accounting for the coordination number smaller than two.
The effect of temperature is to decrease the proportion of two-fold bonded
atoms. This change is clearly correlated with an increase in the number
of one- and three-fold coordinated atoms. This confirms the suggestion
made by many authors, that the effect of temperature is to
decrease the average chain length. However, the high proportion of
three-fold-coordinated atoms
at the highest temperature indicates
that the chains tend to
branch, thus forming an interconnected network. To illustrate this
point, we show in Fig.\ \ref{fig:snap1} and \ref{fig:snap2} `snap-shots'
of typical configurations
at T~=~870 and 1370~K. It is the presence of
a significant number of three-fold-coordinated
atoms at high temperature which is responsible for
the appearance of the shoulder in the pair correlation function
at 2.75~\AA\
and the resulting increase of $g(r)$ at $r_c$.
We point out that, compared with experimental data,
our theoretical pair correlation functions are too large in the region
of the first minimum, particularly for T~=~870~K.
The ratio of $g(r)$ at $r_1$ and $r_c$ is linked to the exchange rate of
atoms within the first coordination shell. The decrease of the ratio with
increasing temperature thus implies a decrease of the average bond
lifetime.

We have used the simulation to study the diffusion of the atoms, by
calculating the time-dependent mean square displacement. From the
slope of this quantity, we estimate the values of
the diffusion coefficients at 870\ K and 1370\ K
to be $0.5\times 10^{-8}$\
m$^2$s$^{-1}$ and $1.5\times 10^{-8}$\ m$^2$s$^{-1}$.
These values are typical of liquid metals, and an interesting question
which further study is how the atoms manage to diffuse so fast in
spite of the well-defined chain structure.

\section{Conclusions}

We have shown that our AIMD simulations give a structure factor for
high-temperature $\ell$-Se that is in excellent agreement with
diffraction data. Analysis of the simulated $g(r)$ gives a
mean coordination number close to 2, but we find
significant fractions of one-fold and three-fold coordinated atoms,
particularly at 1370~K, so that the simple chain-like structure
characteristic of the crystal is considerably disrupted.
The self-diffusion coefficient at both temperatures is high, with
values ($\sim 10^{-8}$~m~s$^{-1}$) typical of liquid metals.

{}~

The work of JMH is supported by EPSRC grant GR/H67935. An allocation
of time on the Fujitsu VPX 240  machine at Manchester Computer Centre
under EPSRC grant GR/J69974 is gratefully acknowledged. Analysis
of the results was performed using distributed hardware provided
under EPSRC grants GR/H31783 and GR/J36266. Discussions with
J.~E.~Enderby and A.~C.~Barnes and technical advice from M.~C.~Payne
played an important role.
We are grateful to Prof. K. Tamura and Dr. S. Hosokawa for sending us
numerical data for their experimental structure factors and radial
distribution functions \cite{tam92}.

\begin{table}
\narrowtext
\setdec 0.0
\caption{Distribution of coordination numbers $N_c$ at T=870 K and
T=1370 K. The cut-off radius $r_c$ is taken to be the first minimum in
$g(r)$, i.e. 2.75 \AA\ and 2.85 \AA\ for the first and the second
temperature respectively. We show the average percentage of atoms
with coordination $N_c$.}
\label{tab:nc}
\begin{tabular}{ccc}
$N_c$       & T=870 K & T=1370 K \\
\tableline
1 & 11.3 & 18.6 \\
2 & 83.7 & 66.9 \\
3 & 4.9  & 13.5 \\
4 & 0.1  & 1.0 \\
\end{tabular}
\end{table}

\begin{figure}
\caption{The total structure factor of liquid Se at 870 and 1370~K.
Solid line represent the calculated $S(k)$.
The full circles
are experimental results from neutron scattering experiments done at
873~K and 1373~K under pressures of 10 and 100~bar, respectively
\protect \cite{edeling81}.
The empty circles are experimental results from X-ray diffraction
experiments done at 873~K and 1473~K under pressures of 16 and 120~bar,
respectively \protect \cite{tam92}.}
\label{fig:sf}
\end{figure}

\begin{figure}
\caption{The radial distribution functions of liquid Se at 870 and 1370~K.
Solid line represent the calculated $g(r)$ and the dotted and the dash-dotted
lines are the experimental curves obtained from X-ray \protect \cite{tam92}
and neutron \protect \cite{edeling81} diffraction experiments, respectively.}
\label{fig:rdf}
\end{figure}

\begin{figure}
\caption{Snapshots of typical configurations of $\ell$-Se
at 870~K.
Bonds are drawn
between Se atoms with separation $<$ 3.0 \AA. Bonds to atoms in neighboring
cells are represented by two-colored sticks.}
\label{fig:snap1}
\end{figure}

\end{document}